\documentclass[aps,showpacs,twocolumn,byrevtex]{revtex4}

\usepackage{psfig,epsfig}
\begin{document}
\title{Observation of near-critical reflection of internal waves\\ in a stably stratified fluid}
\author{Thierry Dauxois}
\email[Email: ]{Thierry.Dauxois@ens-lyon.fr}
 \homepage{http://perso.ens-lyon.fr/thierry.dauxois/}
\author{Anthony Didier}
\author{Eric Falcon}
\email[Email: ]{Eric.Falcon@ens-lyon.fr}
  \homepage{http://perso.ens-lyon.fr/eric.falcon/}
\affiliation{Laboratoire de Physique, {\'E}cole Normale Sup{\'e}rieure de Lyon, UMR-CNRS 5672, 46 all\'{e}e d'Italie, 69007 Lyon, France}
\date{\today}

\begin{abstract}An experimental study is reported of the near-critical reflection of
internal gravity waves over sloping topography in a stratified
fluid.  An overturning instability close to the slope and
triggering the boundary-mixing process is observed and
characterized. These observations are found in good agreement with
a recent nonlinear theory.
\end{abstract}
\maketitle


\section{Introduction}
\label{introduction}

The reflection of near-critical internal waves over sloping
topography plays a crucial role in determining exchanges between
the coastal ocean and the adjacent deep waters. Direct
measurements of mixing in the ocean, using tracers \cite{ledwell12}, have vindicated decades of
phenomenological and theoretical inferences.  In
particular, these measurements have shown that
most of the vertical mixing is not taking place inside the ocean,
but close to the boundaries and topographic features~\cite{polzin97}. These
results have directed attention to the possible role of
internal wave reflection in the boundary-mixing process.

\begin{figure}
\epsfxsize=65mm\epsffile{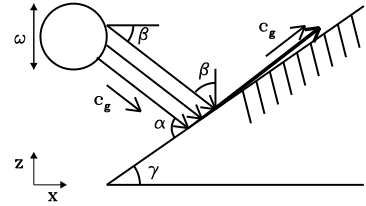}
\caption{Schematic view of the reflection process when the incident
  wave nearly satisfies the critical condition $\gamma \approx \beta$.  The group
  velocity of the reflected wave makes a very shallow angle with the
  slope. $c_g$ indicates the incident and reflected group velocities.
  The reflection law leads to a concentration of the energy density
  into a narrow ray tube.}
\label{schematicreflexion}
\end{figure}

Internal waves have different properties of reflection from a
rigid boundary than do sound or light waves~\cite{phillips}.
Instead of following the familiar Snell's law, internal waves
reflect off a boundary such that the angle with respect to
gravity direction is preserved upon reflection
(Fig.~\ref{schematicreflexion}).  This peculiar reflection law
leads to a concentration of the reflected energy density into a
narrow ray tube upon reflection as displayed in
Fig.~\ref{schematicreflexion}. Theoretical descriptions of this
reflection process have been framed largely in terms of linear and
stationary wave dynamics~\cite{phillips,Gilbert93}.  However, when
the slope angle, $\gamma$, is equal to the incident wave angle,
$\beta$, these restrictions lead to an unrealistic prediction: The
reflected rays lie along the slope with an infinite amplitude and
a vanishing group velocity. Theoretical results have recently
healed this singularity by taking into account the role of
transience and nonlinearity~\cite{JFM}.

Following preliminary oceanographic
measurements~\cite{Sandstom66,Ericksen82},
Eriksen~\cite{eriksen98} has beautifully observed, near the bottom
of a steep flank of a tall North Pacific Ocean seamount, an
internal wave reflection process leading to  a clear departure
from a Garett-Munk model~\cite{garettmunk} for wave frequencies at
which ray and bottom slopes match. Several experimental
facilities~\cite{cacchione,thorpe,ivey,imberger97,maasommeria,Sutherland2000}
have therefore been dedicated to the understanding of the internal
wave reflection and associated instabilities.

However,  results of a controlled laboratory experiment
close to the critical conditions are still lacking since previous ones
with a moving paddle~\cite{cacchione,thorpe,ivey} at one end of a very
long tank, or by the vibration of the tank itself~\cite{maasommeria},
does not generate a clear incident wave-beam as needed for a careful
study. In addition, a direct comparison with the recent and complete
nonlinear theory near the critical reflection would be possible.
Finally, the goal is to improve the understanding of the possible
mixing mechanism near the sloping topography of ocean as very recently
initiated by MacPhee and Kunze~\cite{McPhee02} by exhibiting the
instabilities mechanism leading to mixing. The paper is organized as
follows. The experimental setup is described and carefully explained
in Sec.~\ref{setup}. The main experimental results are presented in
Sec.~\ref{results}, and comparisons with the theory is also provided.
Finally, Sec.~\ref{conclusion} contains conclusions and perspectives.

\section{Experimental Setup}
\label{setup}
The experimental setup consists of a 38 cm long Plexiglas tank, 10 cm
wide, filled up to 22 cm height with a linearly salt-stratified water
obtained by the ``double bucket'' method~\cite{oster}. The choice of
salt, Sodium Nitrate (NaNO$_3$) snow, is motivated due to its highly
solubility in water leading to a salty water viscosity close to the
fresh one. Moreover, this salt allows to reach a strong
stratification: The fluid density ( $1\lesssim \rho(z) \leq 1.2$ g/cm$^3$) measured
at different altitudes ($22\geq z >0$ cm) by a conductimetric probe leads
to a linear vertical density profile of slope $d\rho/d z\simeq -0.0104$
g/cm$^4$.  A rectangular Plexiglas sheet, $3$ mm thick and $9.6$ cm
wide (to allow exchange of water) is introduced at one end of the tank
with an angle $\gamma=35^\circ$ with respect to the horizontal tank bottom (see
Fig.~\ref{schematicreflexion}) to create the reflective sloping
boundary.

Internal waves are generated by a sinusoidal excitation provided
by the vertical motion of a horizontal PVC plunging cylinder
($3.1$ cm in diameter and $9.4$ cm long). The cylinder is located
roughly midway between the base tank and the free surface. This
wavemaker is driven by an electromagnetic vibration exciter powered by a low frequency power
supply. Optical measurements confirm~\cite{anthony} that the cylinder motion is sinusoidal
without distortions for vibrational frequencies $0.2 \leq f
\leq 0.5$ Hz and maximal displacement amplitudes up to $8.5$ mm
(peak to peak).

The well-known and non-intuitive dispersion relation of internal
waves in an incompressible, inviscid and linearly stratified fluid
reads \cite{Lighthill}
\begin{equation}
\omega=\pm N\frac{k_{\perp}}{\mid {\bf k} \mid}=\pm N\sin{\beta}\ {\rm ,}
\label{eq:disp}
\end{equation}
where $k_{\perp}$ is the component of ${\bf k}$ perpendicular to
$z$-axis, $N=\sqrt{-(g/\rho_0)\partial\rho/\partial z}$ is the
constant buoyancy (or Brunt-V{\"a}is{\"a}l{\"a}) frequency,
$\rho(z)$ the fluid density at altitude $z$, $g=981$ cm/s$^2$ the
acceleration of gravity, and $\rho_0 \simeq 1$ g/cm$^3$ a
reference density. Thus, from Eq.\ (\ref{eq:disp}), the wave
frequency, $\omega=2\pi f$, determines the inclination angle
$\beta$ of the phase surfaces with the vertical, and not the
magnitude of the wave vector {\bf k}. From Eq.\ (\ref{eq:disp}),
$\beta$ is also the angle between the group velocity ${\bf
c_g}=\partial \omega/\partial {\bf k}$ and the horizontal, since
${\bf c_g} \perp {\bf k}$. Thus, for  a given stratification and
frequency, internal waves of low (higher) frequency propagate at
low (steeper) angle.

The outward radiation of energy is thus along four beams oriented
at an  angle $\beta$ with the horizontal, the familiar St Andrews
Cross structure~\cite{Mowbraraity}.  The beam propagating directly
toward the slope has been singled out by adding a grid on the
surface of water. This grid strongly damps the three other wave
beams that propagate towards the free surface, and thereby
prevents reflection of such beams. A planar wave pattern
consisting of parallel rays is thus generated from the wavemaker.
Although it is well-known that the spatial spectrum of waves
generated by an oscillating  cylinder is
large~\cite{ApplebyCrighton,Voisin,Sutherland2000}, it has been
experimentally checked that the dominant wavelength is
approximately equal to the cylinder diameter (see below).
Moreover, measurements confirm that the low vibrational amplitudes
of cylinder do not influence strongly the wavelength generated.

Different visualization methods are used to study the reflection
mechanism of such internal gravity waves by a boundary layer.
First, the usual shadowgraph technique~\cite{mer} allows
visualization of the qualitative and global 2D evolution of the
incident and reflected waves. It involves projecting a point
source of light through stratified water onto a screen behind the
tank. The optical refractive index variations induced by the fluid
density variations, allows to observe isodensity lines (or
isopycnals) on the screen, located perpendicularly to the light
source and parallel to the longest wall tank. It is therefore
possible to measure the group velocity angle, $\beta$, and the
phase velocity $v_\varphi$ of the incident wave. For various
frequencies of excitation, $0.2 \leq f \leq 0.5$ Hz, the angle
$\beta$ of the St Andrews Cross is measured on a screen leading to
a linear relation between $\omega$ and $\sin{\beta}$ as predicted
by Eq.\ (\ref{eq:disp}) with a slope of $N=3.1 \pm 0.1$ rad/s, for
the stratified fluid prepared as above. This value is in good
agreement with the above static one obtained from the density profile
with a conductimetric probe. The cut-off frequency is then
$f_c=N/(2\pi)\simeq 0.5$ Hz. By time of flight measurements
between signals delivered by two photodiodes, 1 cm apart, each of
7 mm$^2$ area, located along the propagation direction, a value
$v_\varphi =0.6\pm0.4$ cm/s was obtained for an excitation
frequency $f=0.25$ Hz. When these signals are cross-correlated by
means of an spectrum analyzer, the averaged dephasing time leads
to $v_\varphi=0.7 \pm 0.3$ cm/s, close to the previous value. The
wavelength of the incident wave is thus
$\lambda=v_\varphi/f\simeq3$ cm, corresponding as expected to the
oscillating cylinder's diameter. However, neither this shadowgraph
technique nor one \cite{mer} using passive tracers (fluorescein
dye) is sufficiently sensitive to observe quantitative and local
internal wave properties closed to the reflective boundary layer.

Accordingly, the classical Schlieren method \cite{mer,sutherland}
of visualization has been used. Let us just note that
behind the tank, the light beam is refocused by a 
lens a small distance after a slit (instead of the usual knife
blade to increase contrast) to filter the rays. The slit is
oriented orthogonally to the slope generating straight horizontal
fringe lines in the case of no excitation. The image of the
observation field (strongly dependent of the $7.5$ cm diameter
lens) is focused on the screen by a last lens. The internal wave,
producing density disturbances, causes lines to distort, this
distorting line pattern being recorded by a camera. Note that this
experimental technique is sensitive to the index gradient, and
therefore to the density gradient, {\em orthogonal} to the slit,
i.e. parallel to the slope.

\section{Experimental results and comparison with theory}
\label{results}

\begin{figure*}
\begin{tabular}{cc}
{\bf (a) $t=0$} & {\bf (b) $t=T/5$}\\
\epsfig{file=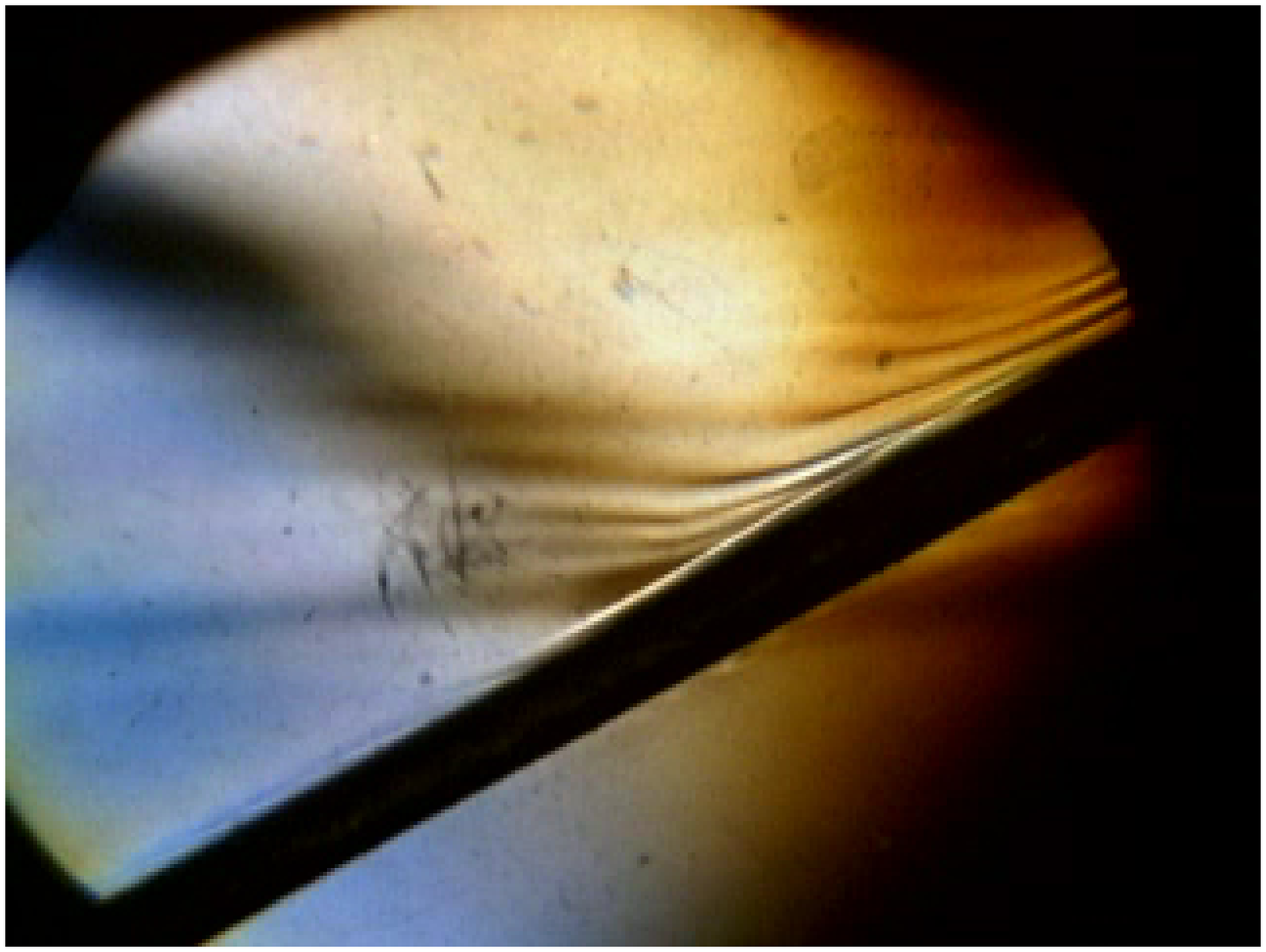, width= 7 cm,  height= 6 cm, angle=180}
&
\epsfig{file=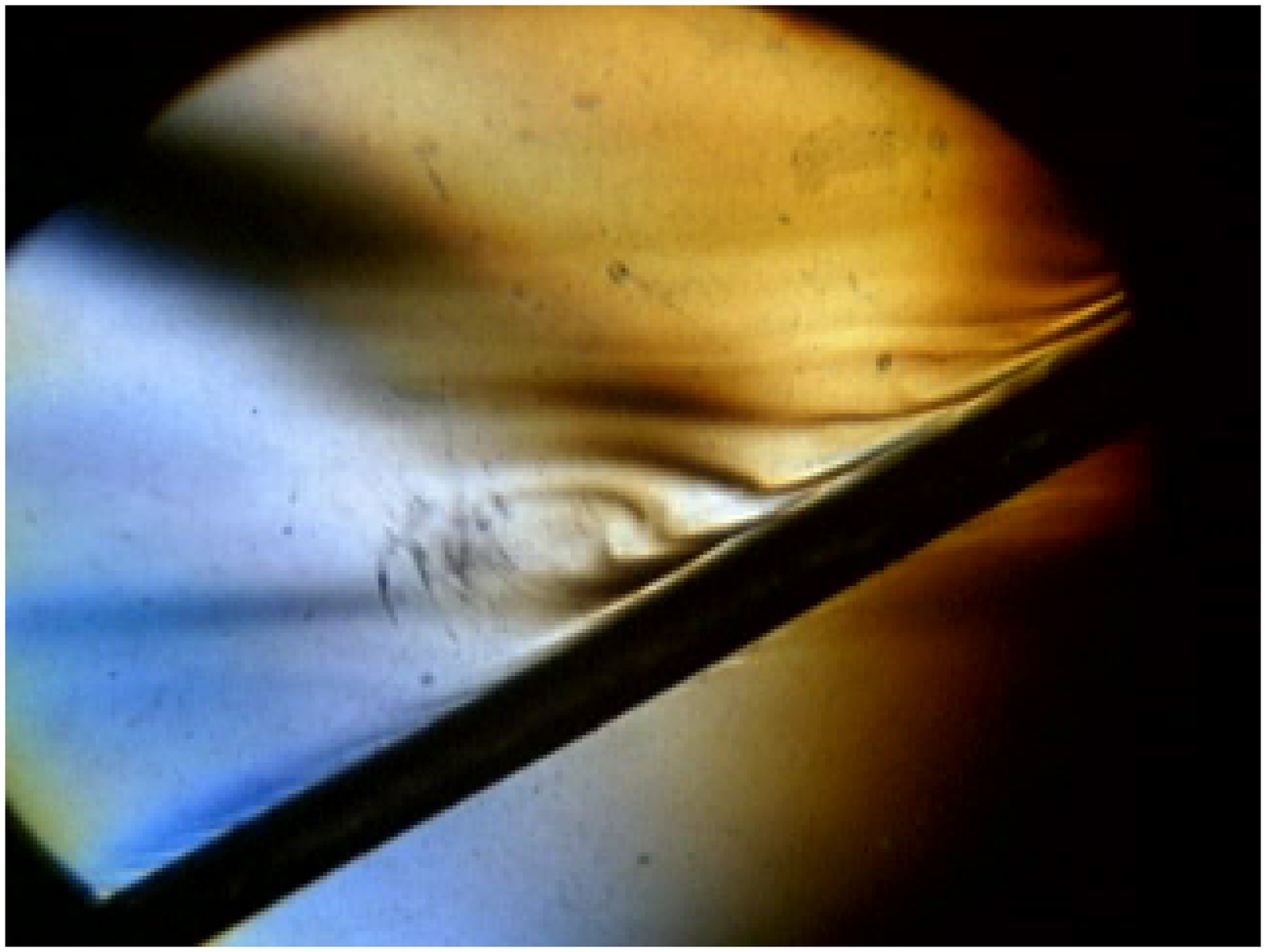, width= 7 cm,  height= 6 cm, angle=180}\\
{\bf (c) $t=2T/5$} & {\bf (d) $t=3T/5$}\\
\epsfig{file=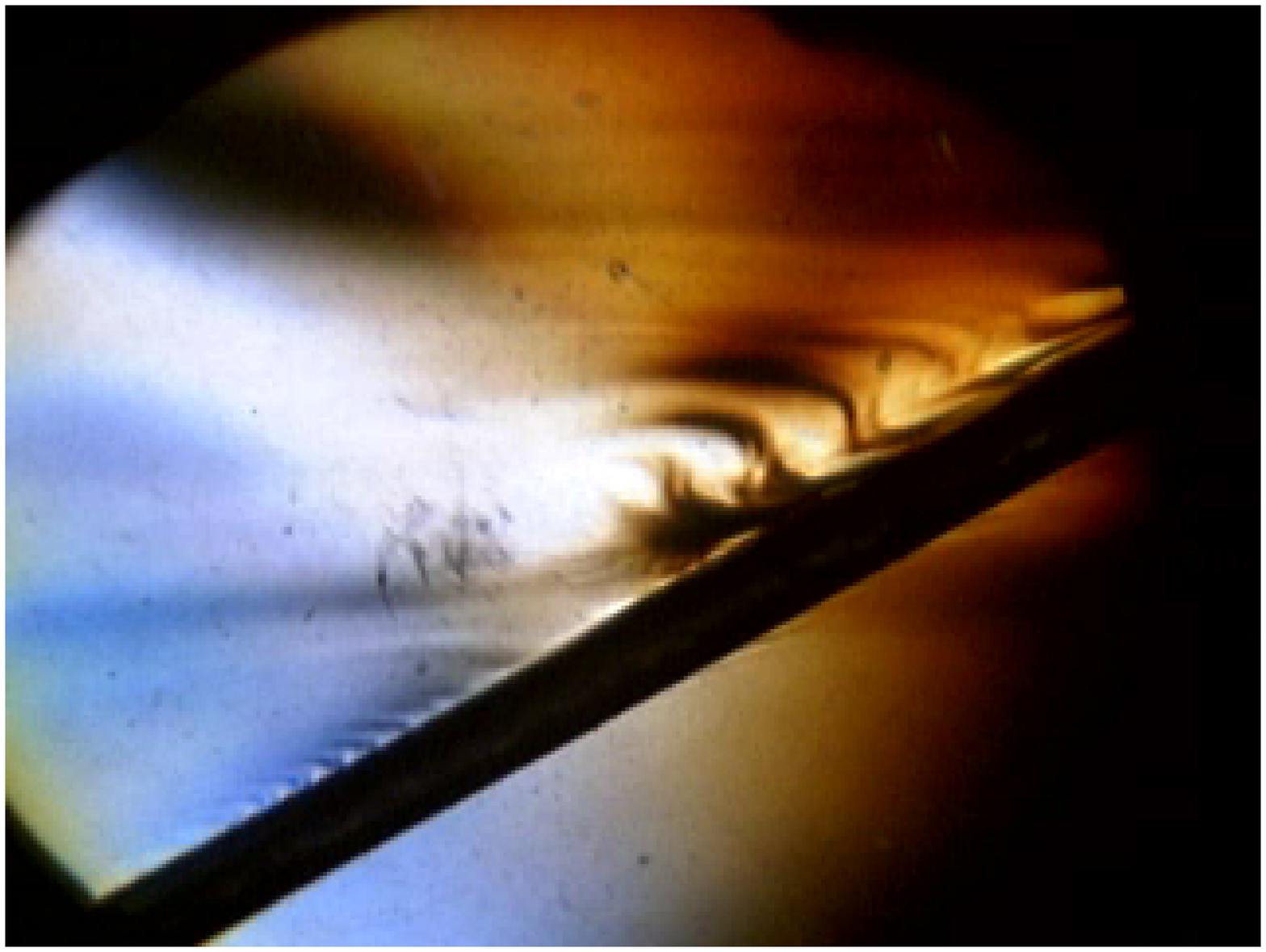, width= 7 cm,  height= 6 cm, angle=180}
&
\epsfig{file=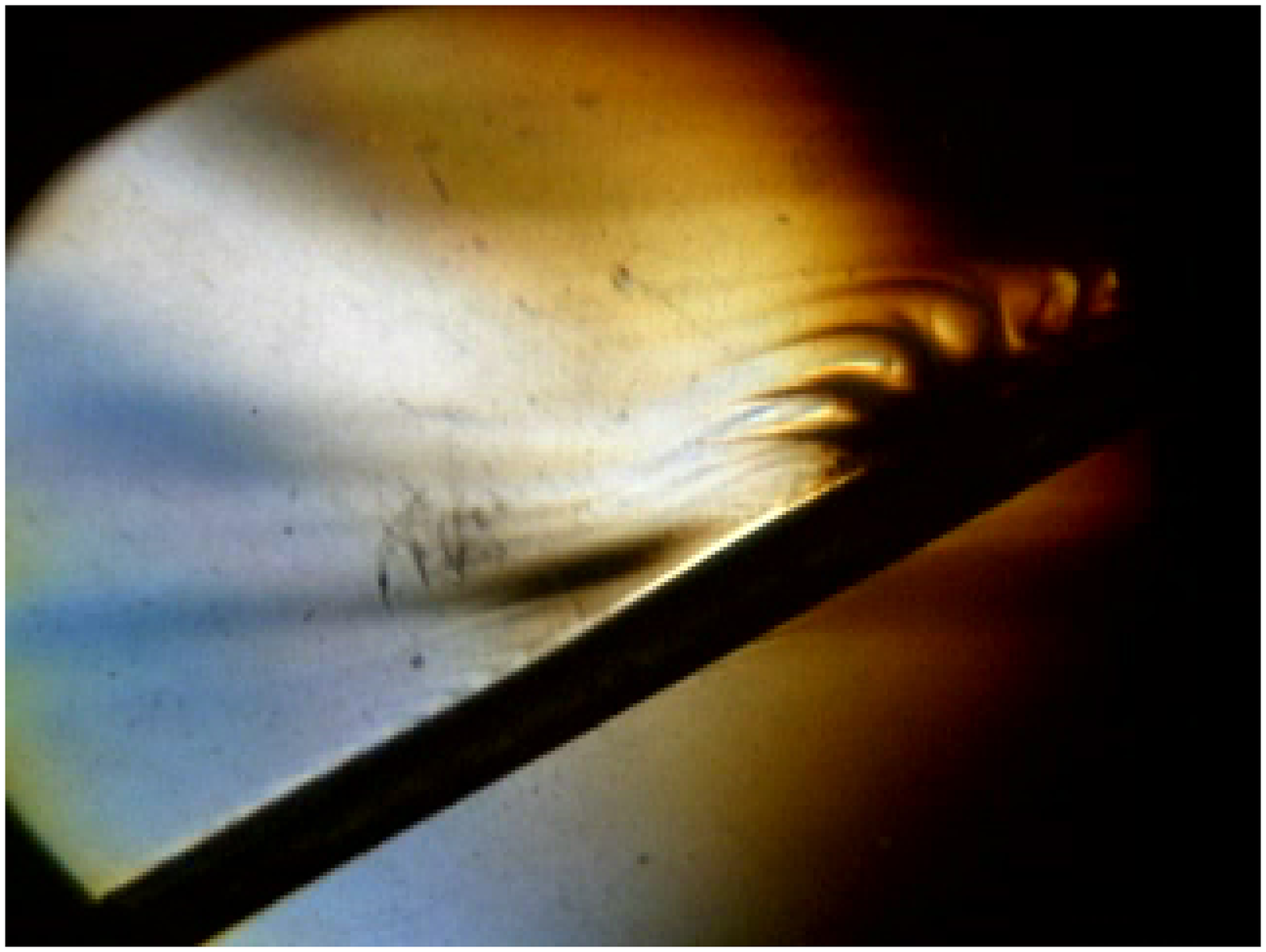, width= 7 cm,  height= 6 cm, angle=180}\\
{\bf (e) $t=4T/5$} & {\bf (f) $t=T$}\\
\epsfig{file=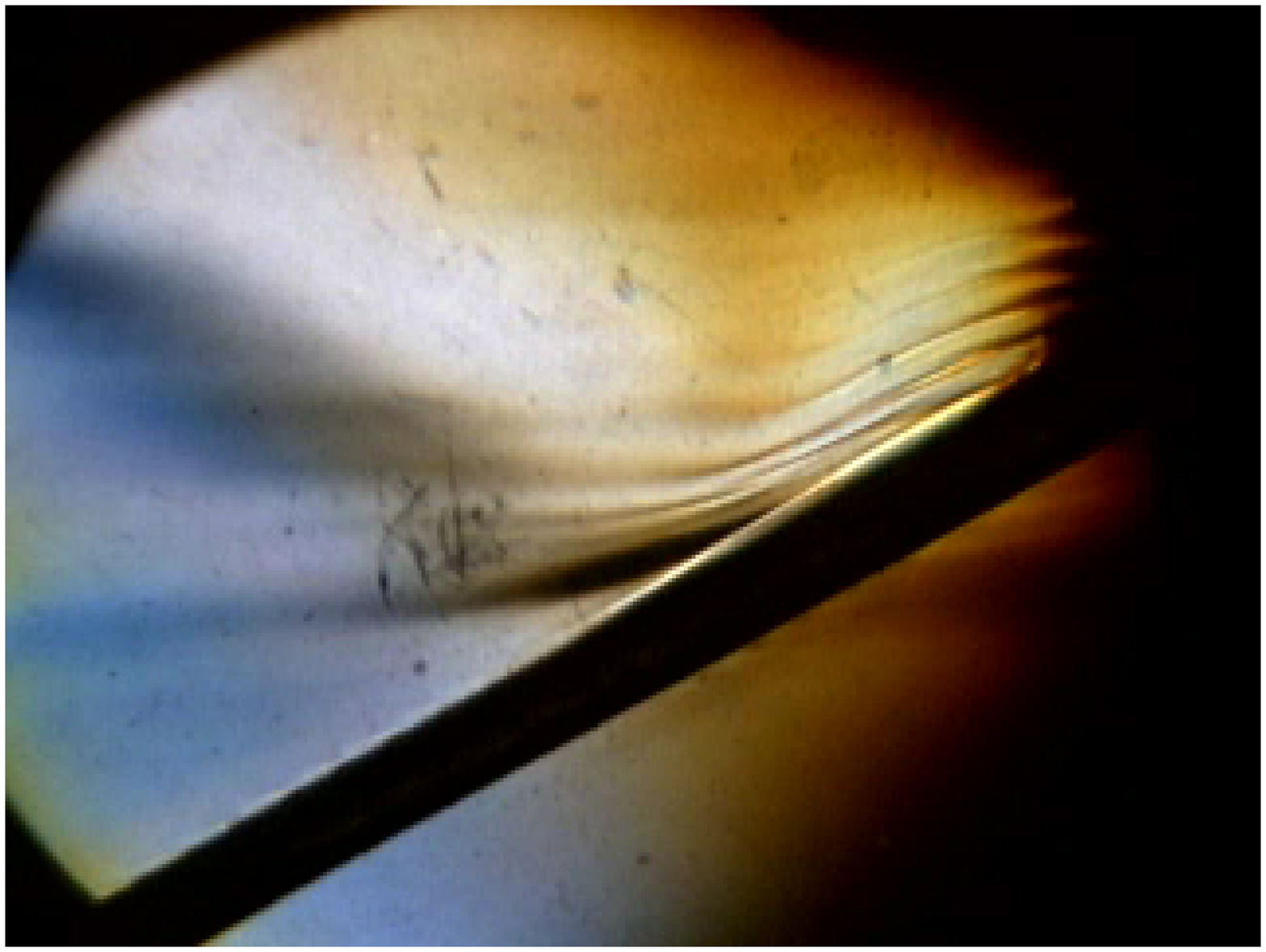, width= 7 cm,  height= 6 cm, angle=180}
&
\epsfig{file=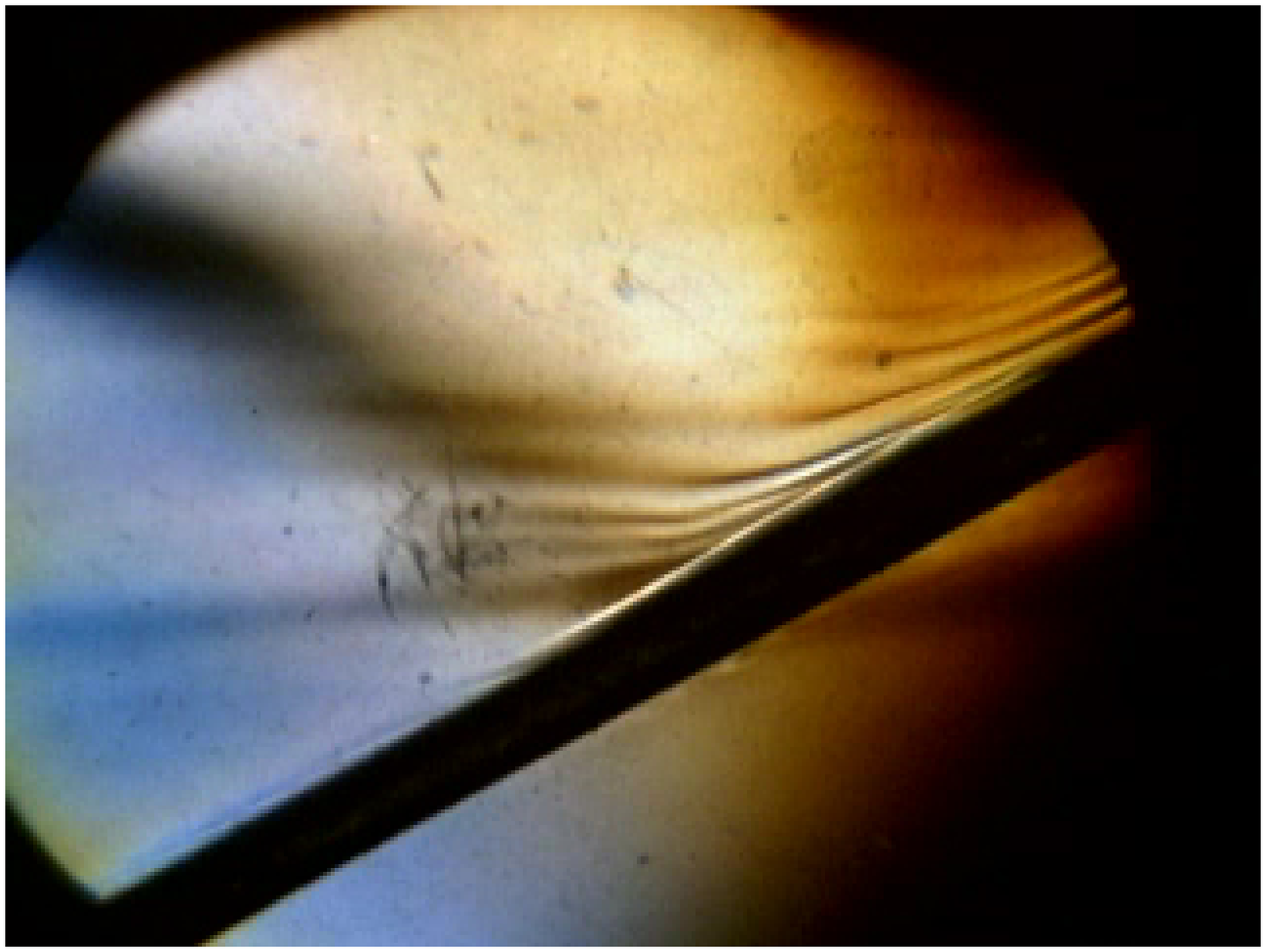, width= 7 cm, height= 6 cm, angle=180}\\
\end{tabular}
\caption{Schlieren pictures showing the slightly subcritical
  reflection ($f/f_c=0.78$) of an internal wave on a slope, during one
  incident wave period $T$. The slope (thick black line) has an angle
  $\gamma=35^\circ$.  The incident wave plane comes in from the left (inclined
  black region near the top left corner between blue and yellow
  regions). The reflected wave plane is hardly noticeable. Wavemaker
  vibrational frequency and peak to peak amplitude are respectively
  $f=0.22$ Hz and 6.7 mm. (Color pictures).}
\label{evol022}
\end{figure*}

The Schlieren technique allows to carefully observe quantitative
and local internal wave properties during the reflection process.
Critical reflection arises when an incident wave beam with an
angle of propagation $\beta$ reflects off the slope of angle
$\gamma \simeq \beta$, the reflected wave being then trapped along
the plane slope. This corresponds to a critical frequency
$f_c=N\sin{\gamma}\simeq 0.28 \pm 0.01$ Hz, $N$ being equal to
$3.1 \pm 0.1$ rad/s for all experiments. It is possible to observe
that the isodensity lines (isopycnals), initially horizontal
without excitation, are bent for an excitation near $f_c$ ($0.78
\leq f/f_c \leq 1.14$), and fold over themselves along the length
of the slope. Figures~\ref{evol022} show a time sequence of
constant density surfaces, depicting sequential snapshots of the
flow throughout one period of its development.
These pictures show strikingly the distortion of the isopycnals in the
slightly subcritical case with $f/f_c=0.78$. In panel~(a), the density
disturbance is very small and one can observe essentially the initial
horizontal background stratification. This background stratification,
usually obtained with dye fluorescein, should be invisible with this
schlieren method.  However, as previously reported by MacPhee and
Kunze~\cite{McPhee02}, the comparison between shadowgraph and Dye
fluorescein visualizations allows to identify these lines with
isopycnals.  In panel (b), the disturbance generated by the incident
wave breaking against the slope begins to `fold-up' the isopycnals. As
time progresses (see panel~c), wave overturning develops around a
front: The buoyancy becomes statically unstable.  This overturned
region climbs along the slope as time continues, and the folded
isopycnals collapse into turbulence that mixes the density field
within the breaking region (see panel~d).  The maximum thickness of
this reflected disturbance is of the order of 5 mm, and decreases with
time as theoretically predicted~\cite{JFM}.  Finally, the flow begins
to relaminarize (panel~e). One can check that panels (a) and (f) are
almost identical, showing that the flow is entirely restratified
(panel~f) after one period of excitation.

\begin{figure}
\epsfig{file=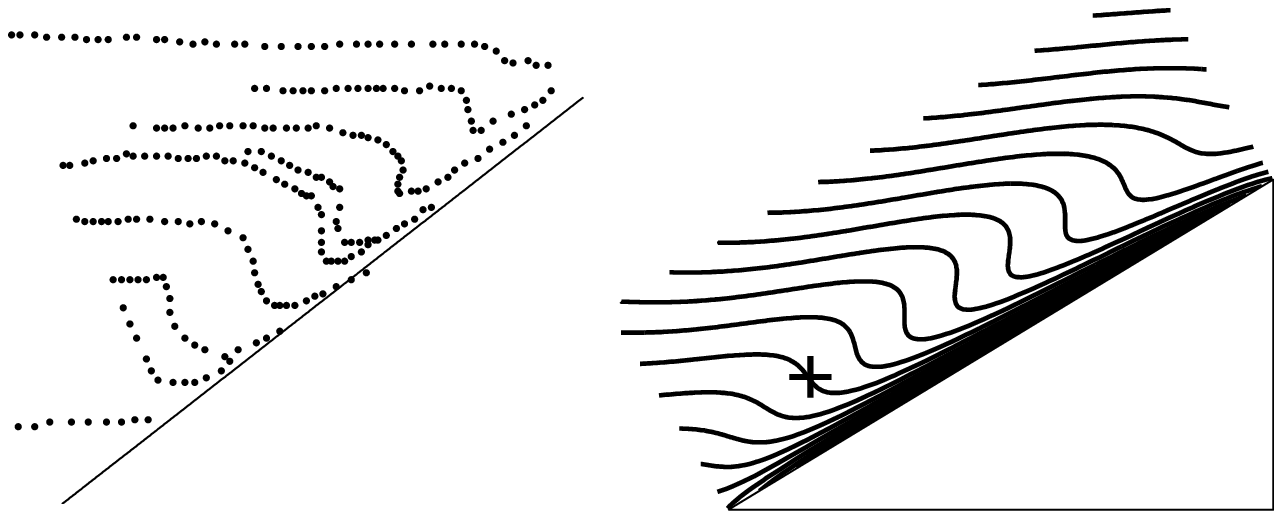, width=8 cm} 
\caption{({\bf a}) Experimental isopycnals extracted from a region of Fig.\
\ref{evol022}c. ({\bf b}) Theoretical isopycnals from Eq.\
(\ref{eq:formrho}). The star indicates the position of the front.
See text for parameters.} \label{evolmat022}
\end{figure}

Figures~\ref{evol022} have been analyzed with an image
processing software (Scion-Image) to extract the isopycnals from
the pictures. A typical experimental result  for the distortion of
isopycnals is reported in Fig.~\ref{evolmat022}a, and is compared
with a theoretical result in Fig.~\ref{evolmat022}b. The analytic
solution of the density field of the initial value problem in the
critical case reads \cite{JFM}
\begin{equation}
  \label{eq:formrho}
  \rho=\rho_0\left[ 1-\frac{N^2}{g}\left(z\cos \gamma-
  \psi B\sqrt{\frac{\mid{\bf k}\mid \sin^2(2\beta)}{2\omega_{+} }}\right)\right]
\end{equation}
where
\begin{eqnarray}
  \label{eq:defB}
  B&=&\sqrt{\frac{ t }{z}}\
  J_1\sin\left(\omega_{+}
    t-\mid{\bf k}\mid \sin(\beta+\gamma) x\right)\quad,\\
 J_1&=&
  J_1\left(2\sqrt{2\omega_{+} \cos^2\beta \ \mid{\bf k}\mid t z}\right)\quad,
\end{eqnarray}
$J_1$ being the Bessel function, $\omega_{+}$ the positive
solution of Eq.\ (\ref{eq:disp}), $\psi$ the maximum amplitude of
the streamfunction and $x$ the horizontal coordinate.
Figure~\ref{evolmat022} show a good qualitative agreement between
experimental and theoretical results, the value for the time $t$
being arbitrarily chosen.  Far from the slope, the density
disturbance is very small and one sees essentially the initial
background stratification. Closer to the slope, the disturbance
folds up the isopycnals, and this produces a region of static
instability.

\begin{figure}
\begin{tabular}{c}
\epsfig{file=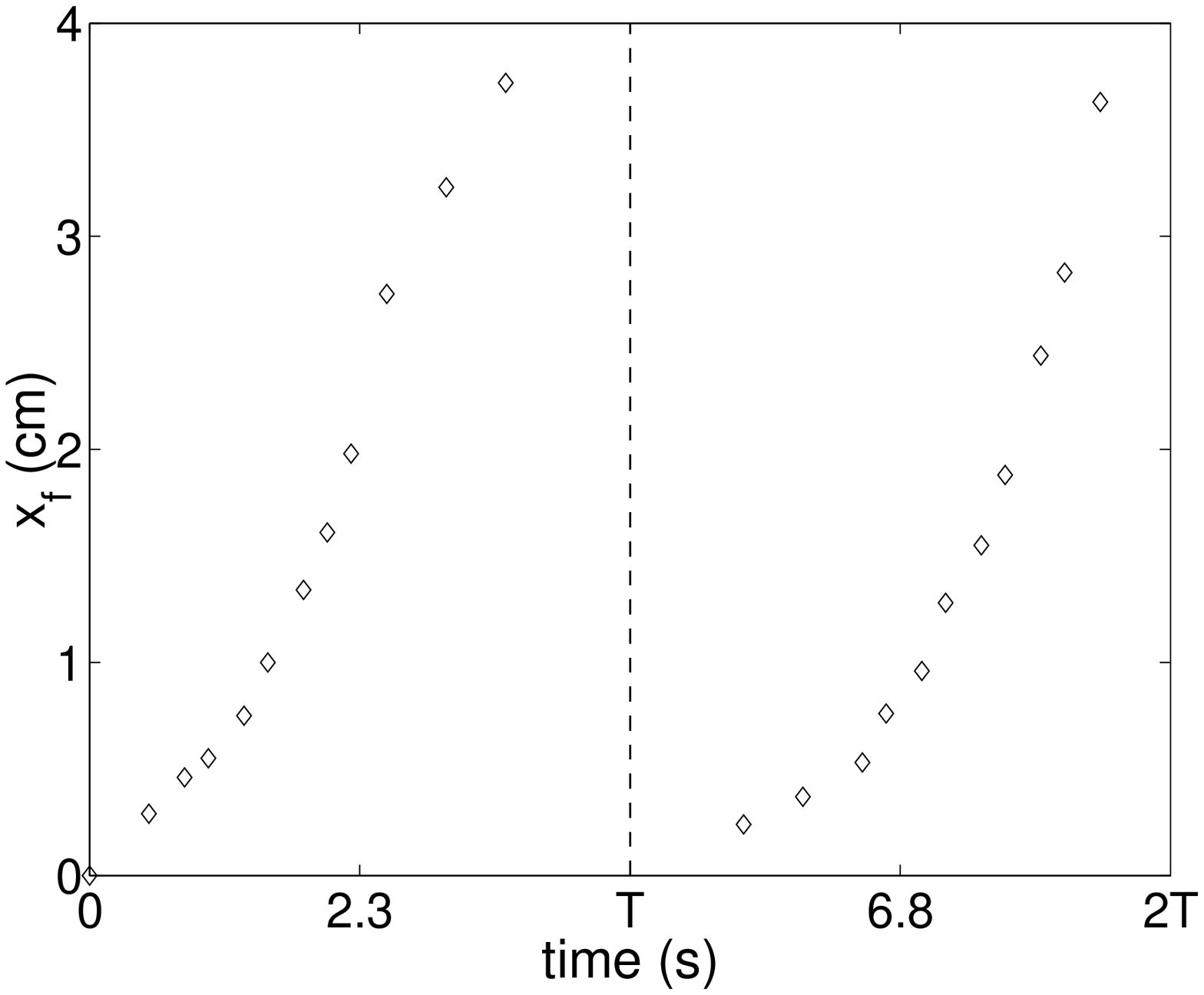, width=6 cm}\\
\epsfig{file=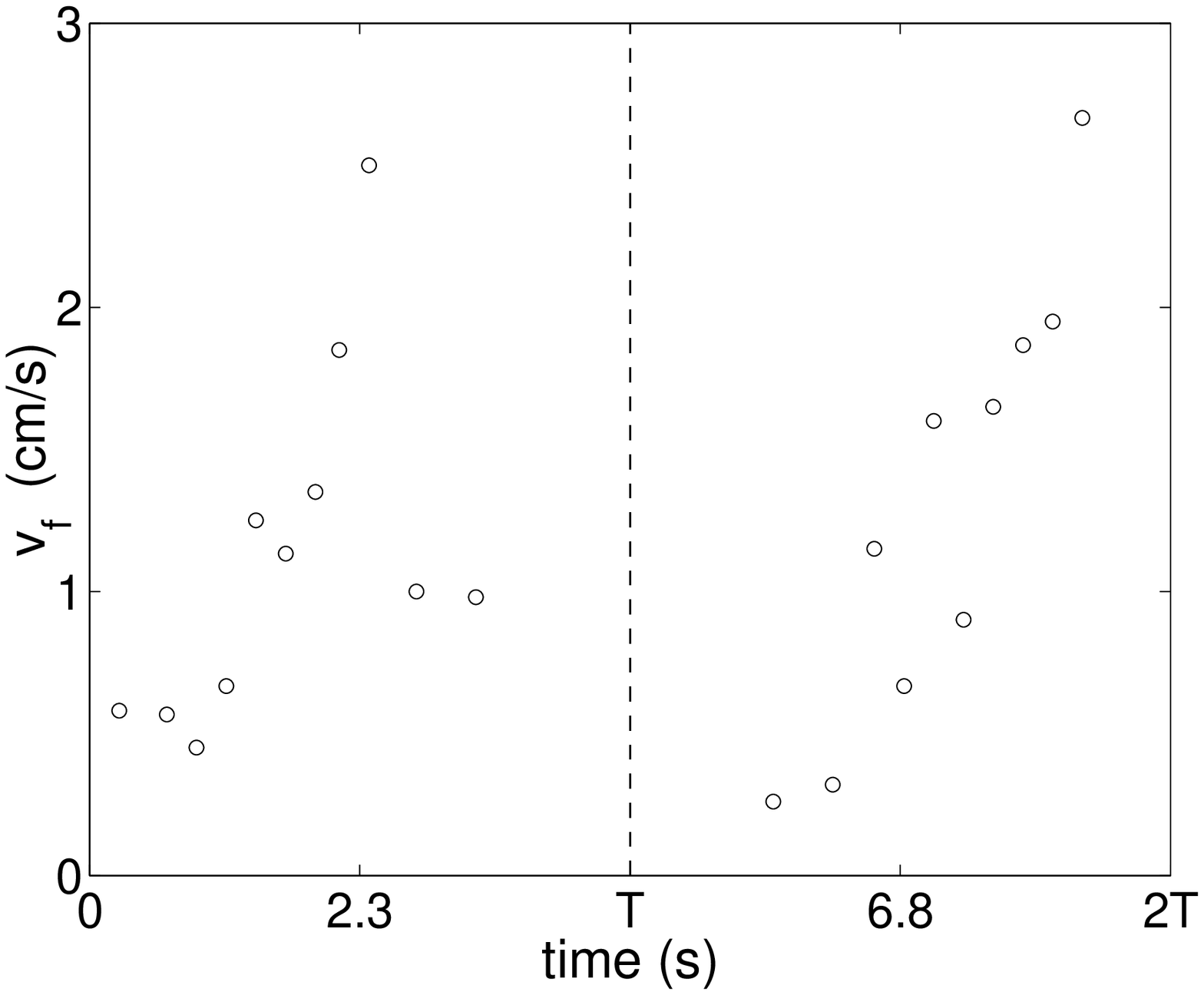, width=6 cm}
\end{tabular}
\caption{Temporal evolution of the isopycnal front position ({\bf
a}) and velocity ({\bf b}) along the slope, during two periods of
vibration. $T\simeq 4.5$ s, $f/f_c=0.78$ and $A_{pp}=6.7$ mm.}
\label{front}
\end{figure}

Recording several isopycnals and using image processing, it is also
possible to follow the temporal evolution of a {\em single isopycnal}
during its overturning. As this phenomenon is periodic with a period
$T=1/f$, it is possible to reconstruct from this temporal evolution
the density profile picture at a given time~$t$.  This allows to
follow the position, and therefore the propagation velocity of the
front along the slope at different times. The front is defined as the
inflexion point (represented by the star in Fig.~\ref{evolmat022}b) of
the followed isopycnal. Figure~\ref{front}a shows, during two periods,
the isopycnal front position along the slope as a function of time.
The periodic evolution of this front position is clearly observed, and
the local slope of curves in Fig.\ \ref{front}a allows to roughly
measure the front velocity as a function of time, as reported in Fig.\ 
\ref{front}b.

The front velocity from its creation to its collapsing increases
from $0.5$ cm/s up to $3$ cm/s. The front has thus traveled
roughly $4$~cm in one period ($\sim4.5$ s). This leads to an
averaged front velocity in agreement with the phase speed
measurement obtained from the shadowgraph visualizations.

\begin{figure}
\epsfig{file=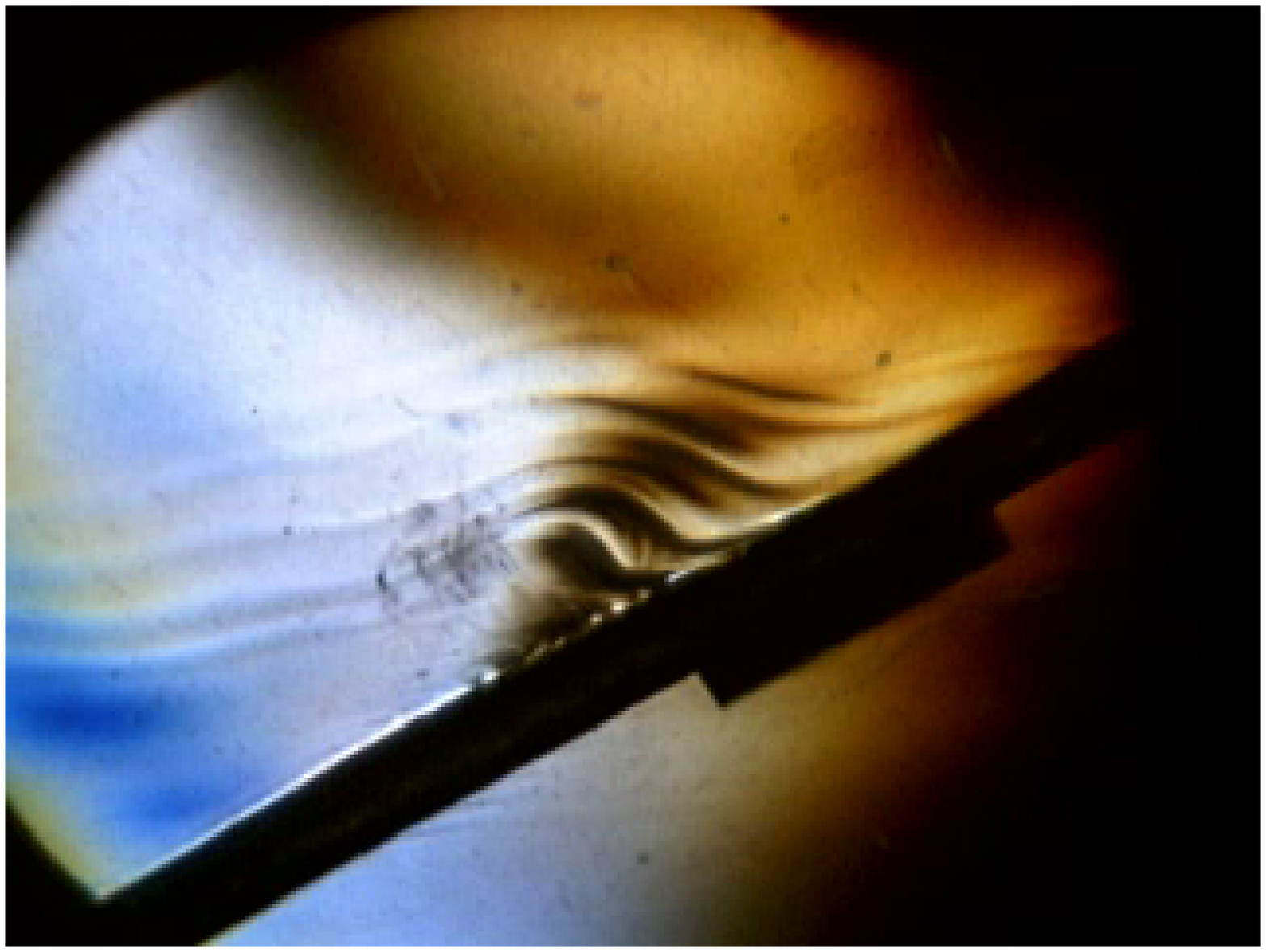, width= 7 cm, height= 6 cm, angle=180}
\caption{Isodensity lines during the slightly supercritical
  reflection ($f/f_c=1.14$) of an internal wave on a slope. The slope
  (thick black line) has an angle $\gamma=35^\circ$. The incident wave plane
  comes in from the left (inclined white region near the top left
  corner between blue and yellow regions). The reflected wave plane is
  hardly noticeable. Vibrational parameters: $f=0.32$ Hz and
  $A_{pp}=6.7$ mm. This picture should be compared with
  Fig.~\ref{evol022}c. (Color picture).}
\label{evol032}
\end{figure}

The wavemaker frequency is now increased up to $f=0.32$ Hz, to
have an incident planewave tilted with an angle~$\beta$ steeper
than the slope angle~$\gamma$.  In this slightly supercritical
case ($f/f_c=1.14$), intrusions are still observed, but the
density field doesn't fold up so abruptly and does not lead
anymore to overturning instability (see Fig.~\ref{evol032}).
Except the value of the frequency~$f$, all others parameter values
have been kept identical to the ones in Fig.~\ref{evol022}. The
instant of this snapshot has been chosen when the isopycnal
distortion is the largest~\cite{webpage}. This distortion is
clearly far from leading to an overturning instability as
encountered in the subcritical case of Fig.~\ref{evol022}c: The
reflected wave is not trapped along the slope in the boundary
layer, and consequently the isopycnals are not overturned.
Moreover the density front velocity is measured roughly constant,
$0.5$ cm/s, during two periods of vibration. Both differences
confirm that the singularity appears only in the critical
case~\cite{JFM}.

\section{Conclusions and perspectives}
\label{conclusion}

The Schlieren technique allows us to study the spatiotemporal
evolution of the internal waves reflection close to the critical
reflection, where nonlinear processes occur. The dynamics of
isopycnals is then found in agreement with a recent nonlinear theory
\cite{JFM}.  Moreover, this experiment confirms the theoretically
predicted scenario for the transition to boundary-layer turbulence
responsible for boundary mixing: The growth of a density perturbation
produces a statically unstable density field which then overturns with
small-scale fluctuations inside~\cite{McPhee02}. Panel~(d) of
Fig.~\ref{evol022} is characteristic of the onset of turbulence
triggered by overturning instability near the slope.

This turbulent mechanism is likely responsible for the formation of
highly ``stepped'' temperature profiles near steep slopes in
lakes~\cite{Caldwell}. Moreover, the formation of suspended sediment
layers, called nepheloid layers, at continental slopes has been linked
to critical angle reflection of internal waves~\cite{white,McPhee02},
and suggests that this reflection process plays an important role in
the seawards transport of sediments in the fluid. A possible extension
of this work, with a smaller slope angle to be closer to real
oceanographic situations, would be to study and characterize the
long time behavior and diffusion process of such particle layers
toward the fluid interior.

\acknowledgments  We warmly thank J.-C.~G{\'e}minard and J.
Sommeria for helpful suggestions.  This work has been partially
supported by the French Minist{\`e}re de la Recherche grant ACI
jeune chercheur-2001 N$^\circ$ 21-31.

\end{document}